# ProtO-RU: An O-RAN Split-7.2 Radio Unit using SDRs

Zhiyu Zhou, Xin Zhe Khooi, Satis Kumar Permal, Mun Choon Chan
National University of Singapore

## ABSTRACT

We present ProtO-RU, the first open source, software-defined O-RAN Split-7.2 Radio Unit built using SDRs and commodity CPUs. Unlike proprietary hardware-based commercial O-RUs, ProtO-RU is built on the open-source srsRAN software stack, and it is fully programmable. We demonstrate that ProtO-RU integrates with the srsRAN and OpenAirInterface5G CU/DU stacks, supports both TDD and FDD duplexing modes, and interoperates with commercial 5G UEs. Our evaluation shows that ProtO-RU remains stable under sustained load with multiple UEs and delivers throughput comparable to Split-8 and commercial O-RUs. ProtO-RU opens up new opportunities for RU-level innovations and lowers the barrier of entry for end-to-end O-RAN research.

## 1 INTRODUCTION

The Open RAN (O-RAN) architecture [29] is one of today's emerging radio access network (RAN) deployment models [8, 32, 37], where the RAN is disaggregated into the Centralized Unit (CU), Distributed Unit (DU), and Radio Unit (RU). In O-RAN, the 7.2 functional split [2, 16] is adopted as it strikes a balanced trade-off between fronthaul bandwidth requirements, RU design complexity, centralized processing flexibility, and the potential for interoperability [5, 27]. These are key to cost-efficient and scalable 5G deployments.

Given O-RAN's popularity, various research efforts have been dedicated to O-RAN-based setups [7, 10, 11, 14, 15, 17, 19, 20, 26, 33, 35, 36, 38–40]. Open-source platforms, such as srsRAN [1] and OpenAirInterface5G [13], have lowered the barrier of entry for researchers to get involved in O-RAN research. However, today's platform only provides researchers with access to the CU and DU, but not the RU. As a result, establishing an end-to-end O-RAN testbed today still requires acquiring commercial O-RUs that support the O-RAN Split-7.2. Fig. 1 summarizes the current state of open-source platforms available for O-RAN research.

Commercial O-RUs are often fixed-function and proprietary black boxes. Their closed design prevents researchers from experimenting with new ideas at the RU, such as extending the Open Fronthaul (OFH) protocol [5], exploring non-standard configurations, evaluating early-stage proposals like shared [21] or multi-numerology RUs, or studying features not yet supported by vendors. Furthermore, COTS RUs often support only a single duplexing mode (TDD or FDD) and a limited range of frequency bands. This inflexibility limits their versatility in diverse research settings.

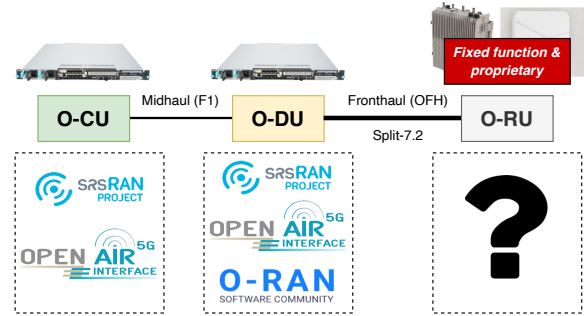

Figure 1: State of open-source platforms for O-RAN research.

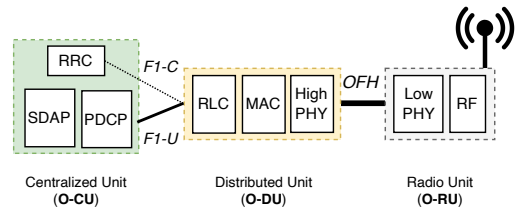

Figure 2: O-RAN architecture and its protocol layers.

Commercial O-RUs aside, software-defined radios (SDRs), such as the USRPs and BladeRFs, are commonly used in research labs as radio frontends for 5G testbed setups. These setups correspond to Split 8 rather than O-RAN Split-7.2. Split-8 setups lack OFH/eCPRI support, do not enforce strict fronthaul timing at the SDR like the O-RU, and push all PHY processing into the DU – unlike the O-RAN Split-7.2, where the PHY is split between the DU and RU (see Fig. 2). This mismatch in capabilities fundamentally limits what researchers can explore or evaluate in the context of O-RAN Split-7.2.

To address the need for an easily accessible, "hackable" and flexible split-7.2 O-RU for O-RAN research, we develop ProtO-RU. ProtO-RU is, to the best of our knowledge, the first open source software implementation of an O-RAN Split-7.2 O-RU. ProtO-RU is built entirely in software, runs on general-purpose processors, and uses SDRs serving as the RF frontends. By extending srsRAN's RU emulator [34] with OFH packet processing support, low-PHY processing, and radio (UHD) integration, ProtO-RU provides a fully functional O-RU that researchers can modify and instrument.

Unlike commercial O-RUs, which rely on FPGAs or ASICs [3, 4], ProtO-RU's software-based implementation allows for fast iteration and customization. And by using SDRs rather than fixed hardware, ProtO-RU naturally supports different duplexing modes and frequency bands, making it highly versatile and well-suited for research environments. As such, ProtO-RU opens up potential new research opportunities, including multi-numerology operation, virtualized and shared O-RUs [21], ML-driven low-PHY processing, OFH security studies [19, 40], and explorations of new OFH protocol designs such as multipoint routing [7].

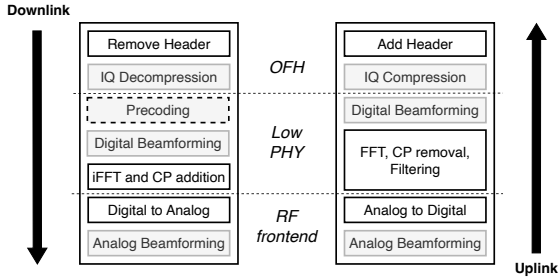

Figure 3: O-RU processing pipeline (adapted from [5]). The shaded blocks are optional features, while precoding (dashed box) is only implemented in Category-B O-RUs.

Using ProtO-RU with USRP B210 and N310 SDRs, we build an end-to-end O-RAN Split-7.2 testbed supporting both TDD and FDD modes. We show that ProtO-RU interoperates with the two major open-source 5G stacks, srsRAN and OpenAirInterface5G (OAI), and operates robustly across long-running experiments. By evaluating ProtO-RU on multiple host platforms, we also demonstrate that its implementation is host-agnostic. Finally, we compare end-to-end latency against a commercial RU to highlight the tradeoffs introduced by a software-based implementation.

**Contribution.** This paper demonstrates that an O-RAN Split 7.2 O-RU can be built entirely in software on general-purpose processors, without using FPGAs or ASICs, something previously believed to be impractical. Through ProtO-RU, we democratize access to the O-RU, lowering the barrier of entry for bringing up an end-to-end O-RAN testbed and engaging in O-RAN research, thereby opening new opportunities for innovation at the O-RU as well as across the broader O-RAN ecosystem.

ProtO-RU is open sourced at https://github.com/NUS-CIR/ProtO-RU, and it is released under Affero General Public License v3.

## 2 BACKGROUND

### 2.1 O-RAN Split-7.2

The 3rd Generation Partnership Project (3GPP) proposed functional splits in the RAN [2, 16], ranging from Split-1 to Split-8. The functional split determines how the different layers of the RAN (e.g., PDCP, RLC, MAC, and PHY) are split across the CU, DU, and RU. We illustrate the case for an O-RAN setup in Fig. 2.

In the context of O-RAN, the 7.2 split is adopted, where the PHY is split into two: high PHY and low PHY, with each residing in the DU and RU, respectively. The DU and RU communicate over the Open Fronthaul (OFH) interface [5], which transports in-phase and quadrature (IQ) samples over an Ethernet-based fronthaul network.

**Open Radio Unit (O-RU).** Fig. 3 illustrates the O-RU processing pipeline, which consists of three major components: OFH handling, low-PHY processing, and the RF frontend. In the downlink pipeline, the RU removes the OFH header, decompresses IQ samples, performs low-PHY operations such as precoding (for Category-B O-RUs), digital beamforming, iFFT, and cyclic prefix (CP) addition, and passes the resulting OFDM symbols to the RF frontend for digital-to-analog conversion and analog beamforming. Uplink processing follows the same sequence in reverse.

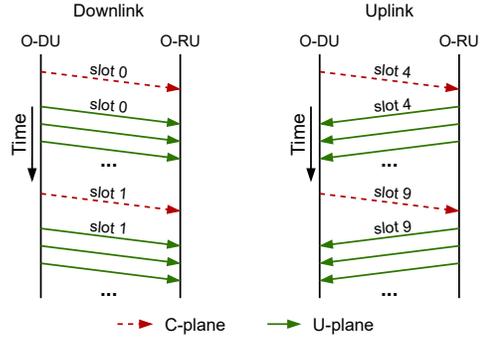

Figure 4: DU-RU message exchange over the OFH interface.

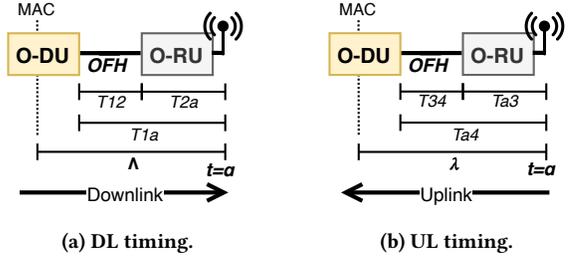

(a) DL timing.  (b) UL timing.

Figure 5: RAN timing in the context of O-RAN Split-7.2.

O-RAN defines two O-RU classes: Category A and Category B. They are differentiated by whether precoding occurs at the RU.

**Open Fronthaul (OFH).** The OFH comprises four logical planes [5]: the control plane (C-plane) and user plane (U-plane) for PHY-layer control and IQ transport, the synchronization plane (S-plane) for time synchronization, and the management plane (M-plane) for RU configuration. In this paper, we focus on the C/U-planes.

Fig. 4 illustrates the typical DU–RU exchange: the DU first sends C-plane packets carrying scheduling information for one or more upcoming OFDM symbols, followed by U-plane packets containing IQ samples. All packets must arrive within a strict timing window. Late arrivals are discarded by the RU.

C-plane packets are generated by the DU for each logical RU antenna port, and they instruct the RU on how to allocate resources to store the incoming IQ sample data in the time-frequency domain. In the time domain, IQ data carried by U-plane packets arrive at the granularity of OFDM symbols, whereas in the frequency domain, one (or more) U-plane packet(s) carry the IQ data for all the physical resource blocks (PRBs).

### 2.2 Timing in the RAN

The RAN enforces strict timing constraints to ensure that data transmissions and receptions occur at precise OFDM symbol boundaries. O-RAN defines several timing parameters, $T1a$ and $Ta4$ for the DU, $T2a$ and $Ta3$ for the RU, and $T12$ and $T34$ for the fronthaul network delay, which are specified at microsecond granularity. Fig. 5 summarizes the timing model for O-RAN Split-7.2.

**Downlink.** As shown in Fig. 5a, the MAC must schedule transmissions ahead of time. For data to be transmitted at $t = \alpha$, the MAC must perform scheduling at $t = \alpha - \Lambda$, where $\Lambda$ is known as the *scheduling offset*. $\Lambda$ must be large enough to account for the high

PHY processing time at the DU, transport delay over the fronthaul network for the IQ samples ($T12$), and low PHY processing time at the RU ($T2a$) for OTA transmission to reach the UE.

**Uplink.** Similarly, in Fig. 5b, an uplink signal received by the RU at $t = \alpha$ only reaches the MAC at $t = \alpha + \lambda$, where $\lambda$ includes the low PHY processing at the RU ($Ta3$), transport delay over the fronthaul network ($T34$), and high PHY processing at the DU.

**OFH timing windows.** For the downlink, the IQ samples must be sent by the DU at $T1a = T12+T2a$, so that the RU can transmit them at the scheduled time. For the uplink, the DU expects IQ samples from the RU to arrive at $Ta4 = Ta3 + T34$.

Given that the fronthaul network delay ($T12$ and $T34$) can vary, the parameters: $T1a$, $T2a$, $Ta3$, $Ta4$, usually come in min/max pairs to define transmission and reception windows that tolerate such variation. Packets that arrive outside of the window are dropped and not processed by the RU. Finally, C-plane packets are always sent ahead of U-plane packets; this advancement is referred to as $Tcp\_adv\_dl$, and it can be up to hundreds of microseconds.

In developing ProtO-RU, one of the key requirements is to determine the timing parameters/ delay profile of ProtO-RU, i.e., $T2a$ and $Ta3$, so that we could derive the $T1a$ and $Ta4$ parameters for the DU – this ensures that IQ samples are sent/received at both the DU and RU at the right time (§3.3).

## 3 PROTO-RU

ProtO-RU builds on the open source srsRAN software stack. Specifically, we extend the existing srsRAN RU emulator [34] by enhancing its OFH library and integrating it with the low PHY and UHD radio libraries. The implementation of ProtO-RU required approximately 6K lines of new code additions/changes[1].

Although ProtO-RU uses existing software libraries, integrating them into a fully functional O-RAN Split-7.2 RU is far from trivial. Below, we outline the key requirements that ProtO-RU must satisfy:

- **Clock synchronization for OTA transmission (§3.1)**: The DU and RU derive fronthaul packet timing from the system clock, which must align with GPS time. However, SDRs operate on independent internal clocks without awareness of system time. Achieving synchronization between these clocks is essential to ensure OTA transmissions occur at the correct slot boundaries.
- **OFH packet handling (§3.2)**: The RU is responsible for handling Open Fronthaul (OFH) packets in real time. This includes processing C-plane packets and dynamically adapting to scheduling instructions from the DU. To accomplish this, the RU must maintain accurate states for downlink, uplink, and PRACH contexts, while ensuring timely resource allocation. Specifically, it must buffer incoming IQ samples for OTA transmission via the low PHY and manage IQ samples received from the low PHY/SDR for forwarding to the DU in the uplink.
- **Determining the delay profile (§3.3)**: To satisfy fronthaul latency constraints, we must derive accurate values for ProtO-RU's $T2a$ and $Ta3$ timing parameters. These determine the DU's corresponding $T1a$ and $Ta4$ values, and ensure that the IQ samples sent/received by the DU are always on time.

---

[1]Based on the diff computed by Git for ProtO-RU's first working release: https://github.com/NUS-CIR/ProtO-RU/commit/35c928b6450b982445026f6f8b4c23e6a2bcc2c1.

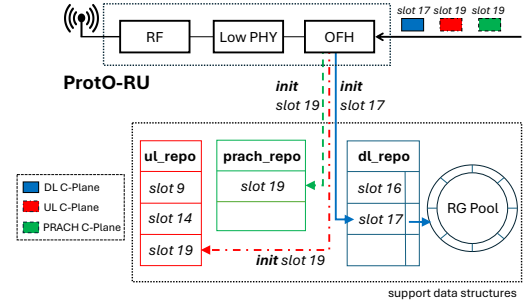

**Figure 6: C-plane packet handling.**

### 3.1 OTA Transmission Time Alignment

First, to support real-time over-the-air (OTA) transmission and reception, the low PHY in ProtO-RU must operate in strict synchronization with actual time, where transmission/reception must happen at precise slot boundaries. However, the low PHY derives its timing from the SDR hardware, which operates on its own internal clock and is unaware of the global time references, i.e., GPS clock.

To address this, ProtO-RU introduces a mechanism to align the hardware-derived time with the GPS clock. It is assumed that the underlying system clock is synchronized with the GPS clock either directly through a GPS receiver or through PTP. During initialization, ProtO-RU queries the current time from the SDR hardware:

```
1  baseband_gateway_timestamp radio_current_time = radio->read_current_time();
```

This value represents the SDR's internal tick count. To ensure that the radio and low PHY begin operation at a clean subframe boundary, ProtO-RU rounds this time up to the next subframe:

```
1  uint64_t sf_duration = static_cast<uint64_t>(cfg.radio_cfg.sampling_rate_hz / 1e3);
2  radio_start_time = divide_ceil(radio_current_time, sf_duration) * sf_duration;
```

The radio and low PHY are then started at this aligned time:

```
1  radio->start(radio_start_time);
2  low_phy->get_controller().start(radio_start_time);
```

Although the offset between the hardware time and the GPS clock is not explicitly labeled in the code, it is implicitly accounted for through this alignment process. This ensures that low PHY operates in the correct slot timing relative to the global time reference.

Finally, timing adapters and notifiers (e.g., `phy_timing_adapter` and `ru_emulator_timing_notifier`) are used to propagate slot boundary events within ProtO-RU.

### 3.2 Open Fronthaul Packet Handling

ProtO-RU extends srsRAN's OFH library to support uplink transmission and adds additional logic to correctly process OFH packets. To ensure low latency, ProtO-RU pre-allocates internal buffers ("repos") and uses flags to track their availability.

OFH packet handling is divided into three flows: (1) C-plane, (2) downlink U-plane, and (3) uplink U-plane.

*3.2.1 C-plane packet handling.* As shown in Fig. 6, ProtO-RU maintains separate repositories for downlink, uplink, and PRACH

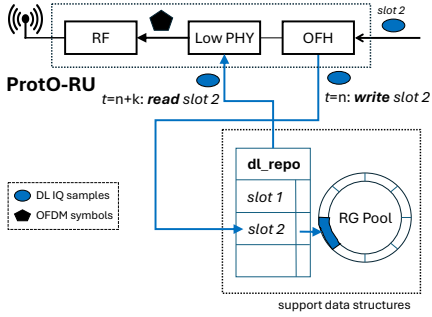

Figure 7: Downlink U-plane packet handling.

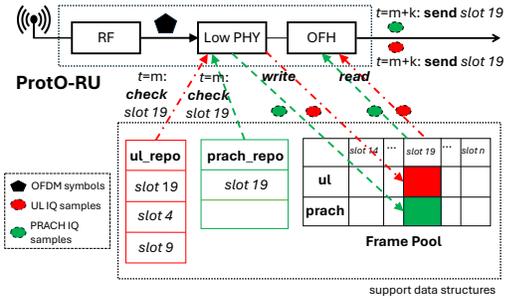

Figure 8: Uplink/PRACH U-plane packet handling.

contexts. Each repository acts as a lookup table for the low PHY, indicating which slots require processing.

When an uplink C-plane packet arrives (e.g., for slot 19) at the OFH, ProtO-RU extracts metadata from the packet header, such as frame number, slot index, symbol IDs, PRB allocation, and section IDs. It then creates an entry in the uplink repository, signaling to the low PHY that IQ samples for that slot are expected. The same logic applies to downlink C-plane and PRACH C-plane packets.

For downlink C-plane handling, ProtO-RU additionally retrieves a resource grid (RG) from a pre-initialized RG pool. The RG provides three-dimensional storage (symbol × subcarrier × antenna port) for downlink IQ samples until low PHY processing.

For PRACH C-plane packets, the frequency offset embedded in the packet is later used by the low PHY to correctly extract PRACH IQ samples in the frequency domain.

*3.2.2 Downlink U-plane packet handling.* When a downlink U-Plane packet arrives at the OFH interface, the RU first extracts the IQ samples from the packet, applies decompression if needed, and then checks whether a corresponding entry exists in the *downlink repository* for the target slot (e.g., slot 2). If so, the IQ samples are written into the RG allocated for that slot; otherwise, the packet is discarded. Later, the low PHY retrieves IQ samples from the corresponding RG for processing (e.g., modulating them into OFDM symbols and cyclic-prefix addition), and then enqueues them for OTA transmission through the radio interface (Fig. 7).

*3.2.3 Uplink U-plane Packet Handling.* As illustrated in Fig. 8, the SDR continuously samples uplink signals and forwards them to the low PHY after analog-to-digital conversion. The low PHY performs standard baseband processing operations such as OFDM demodulation and cyclic prefix removal. After processing, it checks whether a valid entry for the current slot (e.g., slot 19) exists in the *uplink repository*. If so, the IQ samples are encapsulated into OFH-compliant U-plane packets along with the appropriate headers. These packets are then placed in the frame pool and transmitted through the OFH interface at the time defined by $Ta3$.

PRACH samples follow a similar flow but are handled only during PRACH occasions. When a valid entry for a PRACH slot (e.g., slot 19) exists in the *PRACH repository*, the IQ samples are packaged into OFH packets with the `filter_index` marked as PRACH and stored into the frame pool for later transmission by the OFH interface. The low PHY applies the frequency offset provided earlier by the corresponding C-plane packet, ensuring that the PRACH signal extracted is correctly aligned in the frequency domain.

### 3.3 Determining Timing Parameters

To ensure proper scheduling of IQ sample transmission between the DU and RU, we must define ProtO-RU's delay profile, i.e., $T2a$ (downlink) and $Ta3$ (uplink). These determine the timing parameters, $T1a$ (downlink) and $Ta4$ (uplink), used at the DU.

To make the discussion concrete, we assume a subcarrier spacing (SCS) of 30kHz, which corresponds to a slot duration of 500ns. Note that for other SCS, $T2a$ and $Ta3$ will differ.

**T2a.** srsRAN's low PHY processes downlink IQ samples several slots ahead of transmission. Specifically, srsRAN hardcodes this offset to 1.0008ms[2], which translates to 3 slots [12, Section 3.1] (after rounding up). This means that downlink IQ samples must be ready in the RG pool at least three slots in advance.

Additionally, we have to factor in the processing time at the OFH, such as IQ data extraction and decompression. Empirically, this ranges between 1–2 slots. Therefore, using the upper bound, we set ProtO-RU's $T2a$ to approximately five slots.

**Ta3.** For uplink, low PHY processing begins right after RF reception. Based on our empirical measurements, the time taken for the low PHY to produce IQ samples and insert them in the frame pool is approximately 1–2 slots. As such, we ensure that ProtO-RU's $Ta3$ parameter is set to exceed two slots.

*Example.* We show our delay profile for a TDD setup in Table 1; for the delay profile of an FDD setup, please refer to Table 3 in the appendix. Compared to commercial RUs, which usually have a delay profile of less than 1 unit slot [24], ProtO-RU's delay profile is substantially higher, and impacts end-to-end latency performance

Table 1: Delay profile for TDD (SCS = 30 kHz) setup; the DU and RU are directly connected over a 1-meter copper DAC with minimal fronthaul delay.

(a) ProtO-RU delay Profile

| Parameter | in $\mu s$ |
| --- | --- |
| T2a_max_cp_dl | 2635 |
| T2a_min_cp_dl | 2221 |
| T2a_max_cp_ul | 2635 |
| T2a_min_cp_ul | 2221 |
| T2a_max_up | 2454 |
| T2a_min_up | 2015 |
| Ta3_max | 1280 |
| Ta3_min | 925 |

(b) O-DU delay Profile

| Parameter | in $\mu s$ |
| --- | --- |
| T1a_max_cp_dl | 2635 |
| T1a_min_cp_dl | 2335 |
| T1a_max_cp_ul | 2670 |
| T1a_min_cp_ul | 2386 |
| T1a_max_up | 2460 |
| T1a_min_up | 2180 |
| Ta4_max | 1325 |
| Ta4_min | 925 |

---

[2]This corresponds to the variable "`rx_to_tx_max_delay`" in srsRAN.

Table 2: Testbed setup summary.

| | |
|---|---|
| UEs | Quectel RM500Q-GL 5G [30] |
| Commercial O-RU | Benetel RAN550 [6] |
| ProtO-RU (Setup 1) | Intel Xeon Gold 6326; Intel E810 4x25GbE NIC; USRP B210 |
| ProtO-RU (Setup 2) | Intel Xeon Gold 6336; Intel X710 2x10GbE NIC; USRP N310 |
| ProtO-RU (Setup 3) | Intel Core i7-6700; Intel X710 2x10GbE NIC; USRP B210 |
| Time Synchronization | Configuration LLS-C1 (DU as PTP GM) |
| CU/DU Server | Intel Xeon Gold 6433N; Intel E810 4x25GbE NIC |
| Core Network Server | Intel Xeon Gold 6326; NVIDIA CX6 2x100GbE NIC |
| 5G CU/DU stacks | OpenAirInterface5G (tag:2025.w40); srsRAN (tag: version 24.10) |
| 5G Core Network | Open5GS SA Core (v2.7.5) [18] |
| OS/ kernel | Ubuntu 24.04 LTS; Linux kernel v6.8.1-realtime |

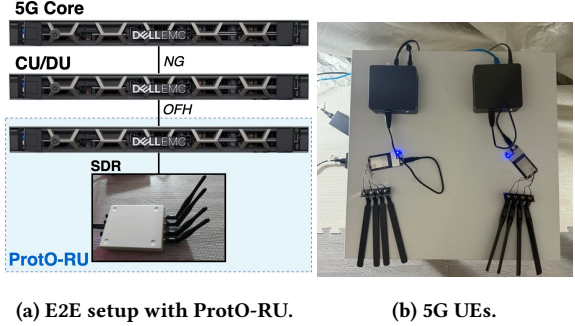

(a) E2E setup with ProtO-RU.    (b) 5G UEs.

Figure 9: Testbed overview.

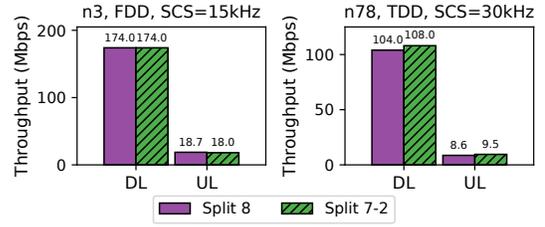

Figure 10: Comparison with Split-8 (20 MHz BW, MIMO 2x2).

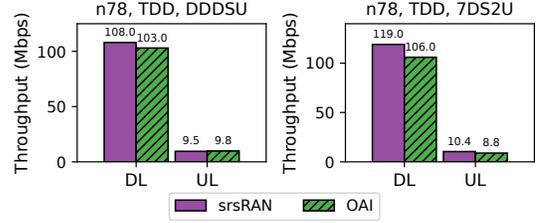

Figure 11: ProtO-RU running with different CU/DU stacks and TDD patterns (20 MHz BW, MIMO 2x2).

(see later in §4). However, this does not affect ProtO-RU's functional operation and stability, as long as the MAC scheduling offset can compensate for the large delay profile.

## 4 EVALUATION

Using ProtO-RU, we set up an end-to-end O-RAN Split-7.2 setup as shown Fig. 9. We also provide a summary of the testbed in Table 2. We report downlink/uplink throughput measured using iPerf3 and RTT measured at the UE using ICMP ping. Unless otherwise specified, all throughput and RTT values are obtained from 30-second iPerf3 runs and 30-second ping sessions with 200-ms intervals.

In our default setup, srsRAN operates as the CU/DU stack, and we use ProtO-RU (Setup 1) in Table 2. As for the default RAN configuration, the bandwidth size is 20 MHz, the MIMO order is 2x2, the MAC scheduling offset is set to 10 (TDD) and 5 (FDD) slots, and the PRACH configuration index is set to 159 (TDD) and 213 (FDD), which corresponds to the short preamble format B4.

### 4.1 Comparison with Split-8

First, we compare the performance of a ProtO-RU-based O-RAN Split-7.2 setup with a baseline Split-8 setup, both of which utilize the USRP B210 SDR as the RF frontend. As in Fig. 10, the measured downlink and uplink throughput is comparable between Split-7.2 and Split-8 under both FDD (n3, 15 kHz SCS) and TDD (n78, 30 kHz SCS) configurations. For example, in the FDD scenario, Split-7.2 achieves 174.0 Mbps DL and 18.0 Mbps UL, while Split-8 achieves 174.0 Mbps DL and 18.7 Mbps UL.

Similarly, RTT performance is nearly identical: 36.99 ms vs. 36.74 ms (FDD) and 32.89 ms vs. 32.87 ms (TDD) for Split-7.2 and Split-8, respectively. These results show that a ProtO-RU-based O-RAN Split-7.2 setup achieves performance comparable to a Split-8 setup.

### 4.2 Interoperability across Different CU/DU Stacks and TDD Patterns

Next, we demonstrate ProtO-RU's interoperability with different CU/DU stacks and TDD patterns. Fig. 11 presents results using srsRAN and OAI under two TDD configurations: DDDSU and DDDDDDDSUU (7DS2U).

Across both stacks and TDD patterns, ProtO-RU maintains similar DL/UL throughput. For example, under DDDSU, we observe 108.0/9.5 Mbps with srsRAN and 103.0/9.8 Mbps with OAI. The RTT results show a similar trend. For DDDSU/7DS2U, we recorded 32.89 ms/31.08 ms and 28.89 ms/29.23 ms for srsRAN and OAI, respectively. These results confirm ProtO-RU's interoperability across different CU/DU stacks and demonstrate that its implementation correctly adapts to varying TDD patterns based on the arrival of C-plane packets.

### 4.3 Multi-UE Long-Running Tests

We evaluate the stability of ProtO-RU under continuous operation using two UEs for over one hour, with concurrent ping (1-second interval) and downlink iPerf3 sessions. To test generality, we repeat the experiment across three different host systems and SDRs (see Table 2) while using different RAN configurations: (1) Setup 1 (FDD, 20 MHz, MIMO 2x2), (2) Setup 2 (TDD, 40 MHz, MIMO 2x2), and (3) Setup 3 (TDD, 20 MHz, SISO).

Across all setups, both throughput and RTT remain stable and uninterrupted throughout the hour-long tests (see Fig. 12). Higher-end systems (Fig. 12a and 12b) exhibit more stable performance throughout the experiment[3], while the lower-end setup (Fig. 12c) experiences slightly more packet drops (marked in red) and throughput fluctuations. This is well within acceptable limits for experimentation and testbed use. These results indicate that ProtO-RU is robust

---
[3]For setup 2 (Fig. 12b), we note that the throughput performance remains far from the theoretical maximum. This indicates that the RAN's configuration requires further tuning, although the current BLER remains low at typical ranges.

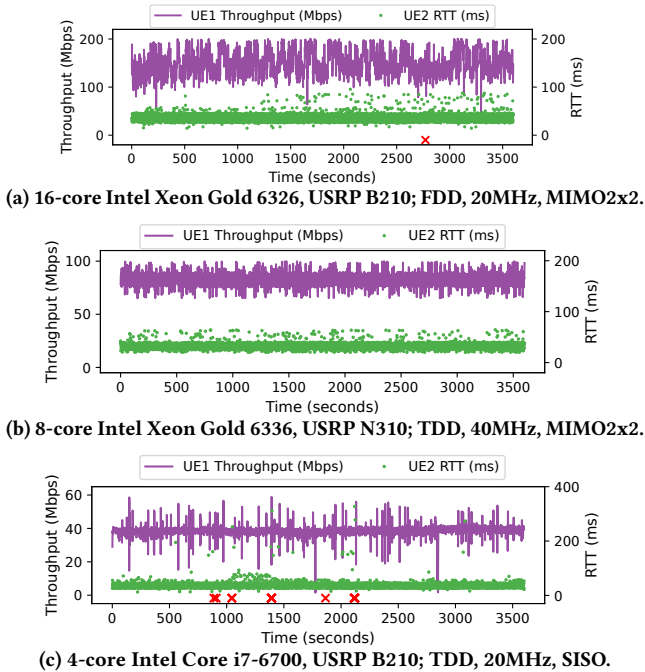

(a) 16-core Intel Xeon Gold 6326, USRP B210; FDD, 20MHz, MIMO2x2.

(b) 8-core Intel Xeon Gold 6336, USRP N310; TDD, 40MHz, MIMO2x2.

(c) 4-core Intel Core i7-6700, USRP B210; TDD, 20MHz, SISO.

Figure 12: Multi-UE long running tests with ProtO-RU running on different host systems and SDRs.

enough for end-to-end experimentation, suitable for prototyping, and even for testbed deployments [9].

### 4.4 Comparison with Commercial O-RU

Finally, we compare ProtO-RU against a commercial O-RU, a Benetel RAN550 RU [6]. For fair comparison, we configure the commercial O-RU to operate at TDD, 40 MHz, MIMO 2x2, and compare it with ProtO-RU (Setup 3 of Table 2, and Fig. 12b). We focus on the RTT measurements given their differences in their delay profiles.

Across 5000 samples collected from 1-second interval pings, we note that when using the RAN550, the average, median, and $99^{th}$-percentile RTT are 16.5 ms, 17.7 ms, and 55.1 ms, respectively. As for ProtO-RU, the RTT is much higher, with the average, median, and $99^{th}$-percentile being 32.9 ms, 32.3 ms, and 59.9 ms, respectively. The high RTT when using ProtO-RU is expected, given the larger delay profile, as more time is needed for processing.

## 5 DISCUSSION

**Fundamental limitations.** While ProtO-RU demonstrates that a fully functional Split 7.2 O-RU is feasible, its software-based processing pipeline introduces a significantly larger delay profile than FPGA/ASIC-based commercial O-RUs. This results in higher end-to-end latency, making ProtO-RU unsuitable for latency-sensitive use cases such as URLLC. In addition, the scalability of ProtO-RU is limited by the computational capabilities of general-purpose CPUs, and the RF bandwidth and MIMO constraints of commodity SDRs impose practical limits. Nevertheless, ProtO-RU remains a valuable platform for research and prototyping, providing flexibility and programmability unavailable in commercial O-RUs.

**Current features support, and future work.** At the time of writing, ProtO-RU currently implements all mandatory features (Fig. 3) defined in the O-RAN Category A specifications [5]; Category B features such as precoding, beamforming, and massive MIMO are not yet supported, and M-plane functionality for remote configuration is currently absent. ProtO-RU supports BFP compression and has been evaluated only within FR1 (sub-6 GHz), with 15 and 30 kHz SCS, and up to a 2x2 MIMO configuration. For the C-plane, the current implementation assumes per-slot C-plane packets. For PRACH, only short preambles with format B4 are supported.

Moving forward, we plan to expand ProtO-RU's feature set, including broader PRACH format support and Category B functionality. We also intend to optimize the processing pipeline following the findings in [12] to reduce ProtO-RU delay profile. In addition, future iterations may explore the use of general-purpose accelerators (e.g., GPUs) or inline heterogeneous compute engines to improve processing efficiency and support larger MIMO orders.

**Long-term sustainability.** Following email exchanges with the srsRAN project maintainers, we plan to upstream ProtO-RU in the near future, so that it can benefit from broader community support and ensure the long-term sustainability of ProtO-RU.

## 6 RELATED WORK

**O-RU emulators.** Several existing tools emulate O-RUs for testing fronthaul behavior. The srsRAN RU Emulator [34], O-RAN Software Community xRAN-based O-RU sample app [23], and NVIDIA Aerial RU Emulator [22] validate DU-generated fronthaul timing and generate uplink U-plane packets in response to C-plane messages. Neither implements the low PHY processing pipeline nor integrates an RF frontend, and therefore cannot support end-to-end OTA communication with actual UEs. In contrast, ProtO-RU extends the srsRAN RU emulator with real-time OFH handling, low PHY processing, and SDR integration; it is the only fully functional O-RU that supports E2E OTA communication with real UEs.

**O-RU hardware platforms.** Recent industry efforts have focused on leveraging high-end SDR platforms [25] to develop Split-7.2 RUs by implementing the OFH interface and the low-PHY layer on the SDRs' on-board FPGAs. In parallel, academic research has explored the design of high-performance O-RUs [28, 31], targeting real-world deployments while also providing open platforms for prototyping and experimentation. Different from these platforms, which either require high-end SDRs or FPGAs, ProtO-RU is entirely software-based, runs on general-purpose processors, and can operate with low-cost SDRs as the RF frontend, significantly improving accessibility and lowering the barrier to entry.

## 7 CONCLUSION

In ProtO-RU, we show that an O-RAN Split 7.2 O-RU can be built using software running commodity CPUs and SDRs. ProtO-RU is open sourced at https://github.com/NUS-CIR/ProtO-RU. ProtO-RU offers a programmable platform and lowers the barrier to building end-to-end O-RAN testbeds. This opens up opportunities for innovation at the O-RU and is key to making O-RAN research even more accessible. Lastly, we invite the community to contribute, extend, and help shape the future evolution of ProtO-RU as a shared open research platform for advancing O-RAN.

# A APPENDIX
## A.1 Setting up ProtO-RU
This description below is to give a general overview of the setup and system requirements of ProtO-RU. For more detailed information, please refer to ProtO-RU's GitHub repository at https://github.com/NUS-CIR/ProtO-RU/.

*A.1.1 System requirements.* To run ProtO-RU reliably, several system requirements must be satisfied to meet the real-time performance and functionality needs:
- CPU:
  – At least 4 cores @ ≥ 2.0 GHz.
  – Higher core counts recommended for larger bandwidths.
- Memory:
  – Minimum 8 GB RAM
- Fronthaul NIC:
  – A 10 GbE (or above) NIC is required.
  – NIC must support hardware PTP timestamping.
- SDR hardware:
  – NI USRPs (only SDRs supported by the srsRAN RU library).
  – Other SDRs (e.g., BladeRF) may work, but have not been tested.

*A.1.2 Testbed setup.* ProtO-RU can either run on a separate server or on the same server as the DU. Depending on the testbed setup topology, e.g., if over two different servers like Fig. 9, the DU and ProtO-RU host system(s) must be time synchronized over PTP. We have prepared a quick guide on setting up PTP time synchronization in our GitHub repository's documentation.

*A.1.3 Installing dependencies.* Since ProtO-RU is based on the srsRAN project, it inherits all the dependencies of srsRAN. Please refer to the srsRAN project documentation, sections "Build Tools and Dependencies" and "RF-drivers" (especially UHD drivers) for the list of dependencies required to build ProtO-RU. If your current system can successfully build and run srsRAN in Split-8 mode with any UHD-compatible SDR (e.g., B210, N310), then you should be able to build and run ProtO-RU without any issues.

*A.1.4 Building ProtO-RU.* To download and build ProtO-RU, first, clone the ProtO-RU repository:

```
git clone https://github.com/NUS-CIR/ProtO-RU.git
```

Then, build ProtO-RU.

```
cd ProtO-RU
mkdir build
cd build
cmake ../
cd ./apps/examples/ofh/
make -j $(nproc)
```

*A.1.5 Running ProtO-RU.* To run ProtO-RU, run the following command from the "build/apps/examples/ofh/" directory (assuming the configuration file is located at "/path/to/ru_emu.yml"):

```
sudo ./ru_emulator -c /path/to/ru_emu.yml
```

**Important Note:** Given that ProtO-RU requires a large delay profile, the integration with other 5G CU/DU implementations may require some patching. We have provided sample working configuration files for both the gNB/DU and ProtO-RU in our GitHub repository located under the "./proto-ru/conf-files" directory.

*A.1.6 Troubleshooting ProtO-RU.* Details on some of the common issues encountered when running ProtO-RU and the recommended remediation steps are described in our GitHub repository, under the file "/proto-ru/TROUBLESHOOTING.md".

## A.2 FDD Delay Profile
In Table 3, we show the delay profile used for our FDD setup.

**Table 3: Delay profile for FDD (SCS = 15 kHz) setup; the DU and RU are directly connected over a 1-meter copper DAC with minimal fronthaul delay.**

(a) ProtO-RU delay Profile

| Parameter | in $\mu s$ |
| --- | --- |
| T2a_max_cp_dl | 4135 |
| T2a_min_cp_dl | 3721 |
| T2a_max_cp_ul | 4135 |
| T2a_min_cp_ul | 3721 |
| T2a_max_up | 3954 |
| T2a_min_up | 3515 |
| Ta3_max | 1480 |
| Ta3_min | 1125 |

(b) O-DU delay Profile

| Parameter | in $\mu s$ |
| --- | --- |
| T1a_max_cp_dl | 4135 |
| T1a_min_cp_dl | 3886 |
| T1a_max_cp_ul | 4135 |
| T1a_min_cp_ul | 3886 |
| T1a_max_up | 3990 |
| T1a_min_up | 3680 |
| Ta4_max | 1500 |
| Ta4_min | 1125 |

## A.3 Time Alignment
Here, we expand on the process of time alignment mentioned in §3.1 about how low PHY calculates the two slot points from both "radio_start_time" and GPS clock, as well as the calculation of the offset between these two slot points.

*A.3.1 Derivation of slot points.* The complete code body can be found in the file: "srsRAN/lib/phy/lower/lower_phy_baseband_processor.cpp".

*Slot point from hardware.* The "start" function first adds a delay "rx_to_tx_max_delay"(the value of which is explained in §3.3) to the "radio_start_time"(mentioned in §3.1) to get the actual start time for downlink processor:

```
void lower_phy_baseband_processor::start(
    baseband_gateway_timestamp init_time)
{
    baseband_gateway_timestamp start_time = init_time +
        rx_to_tx_max_delay;
    ...
```

It calculates the slot point from the actual start time. First, the index of the subframe within the hyperframe (containing the maximum number of frames) is derived, and "nof_samples_per_subframe" can be known from the configured sampling rate from the configuration file. The variables named in capital letters are fixed global values:

```
auto i_sf =
    static_cast<unsigned>((start_time /
        nof_samples_per_subframe) % (NOF_SFNS *
        NOF_SUBFRAMES_PER_FRAME));
```

Then the sample index within the subframe is calculated:

```
unsigned i_sample_sf = start_time %
    nof_samples_per_subframe;
```

And with the sample index we get, the symbol index within the subframe("i_symbol_sf") and the sample index within the symbol("i_sample_symbol") can both be derived. The array "symbol_sizes" represents the number of samples each symbol contains:

```
1  unsigned i_sample_symbol = i_sample_sf;
2  unsigned i_symbol_sf     = 0;
3  while (i_sample_symbol >= symbol_sizes[i_symbol_sf]) {
4      i_sample_symbol -= symbol_sizes[i_symbol_sf];
5      ++i_symbol_sf;
6  }
```

Finally, the slot number from the SDR hardware:

```
1  unsigned i_slot    = i_sf * nof_slots_per_subframe +
       i_symbol_sf / nof_symbols_per_slot;
```

*GPS clock slot point.* The calculation of the slot point from the GPS clock is easier to comprehend since the clock tells the actual time in nanoseconds. "gps_clock" is a class we integrate into the library based on srsRAN implementation. We first get the current time by calling the method "now":

```
1  auto now = gps_clock::now();
```

Then we get the nanosecond fraction of the current timestamp:

```
1  auto ns_fraction = gps_clock::calculate_ns_fraction_from(
       now);
```

The integral part decides the subframe index within a frame, and the fractional part decides the slot index within the subframe.

Finally, the slot point from the GPS clock is calculated, where "scs" represents subcarrier spacing:

```
1  slot_point gps_slot = gps_clock::calculate_slot_point(scs
       ,
2      std::chrono::time_point_cast<std::chrono::seconds>(
           now).time_since_epoch().count(),
3      std::chrono::duration_cast<std::chrono::microseconds
           >(ns_fraction).count(),
4      1000 / get_nof_slots_per_subframe(scs));
```

*A.3.2 Offset calculation.* Once the two slot points are known, the difference between them can be easily calculated by subtraction:

```
1  uint32_t diff = calculate_slot_diff(phy_slot, gps_slot);
```

The definition of function "calculate_slot_diff" is as follows:

```
1  uint32_t calculate_slot_diff(slot_point src, slot_point
       dst)
2  {
3    int diff = dst.system_slot() - src.system_slot();
4    int dis = diff >= 0? diff : diff+dst.
         nof_slots_per_system_frame();
5    return dis;
6  }
```

We assume the hardware slot is always running behind the slot from the GPS clock, so there should be a wrap-around, as shown in line 4. Then, we add the time advancement introduced by "rx_to_tx_max_delay" to get the final offset:

```
1  unsigned cali =   ceil(static_cast<double>(
       rx_to_tx_max_delay)/static_cast<double>(
       nof_samples_per_subframe/nof_slots_per_subframe));
2  diff += cali;
```

This offset is carried all the way through the low PHY processing to make sure the low PHY can map the data to the correct slot number.